\documentclass[
    aps,
    prb,
    twocolumn,
    10pt,
    superscriptaddress,
    floatfix
]{revtex4-1}

\usepackage{graphicx}
\usepackage{amsmath}
\usepackage{amssymb}

\begin{document}

\title{Electronic, structural and magnetic properties of goldene nanoribbons}

\author{Fabr\'icio M. de Vasconcelos}\email{fabricio.vasconcelos@ifpi.edu.br}
\affiliation{Instituto Federal do Piau\'i - Campus Piripiri, CEP 64260-000, Piripiri, Piau\'i, Brasil}
\author{Kleuton A. L. Lima}
\affiliation{Department of Applied Physics and Center for Computational Engineering and Sciences, State University of Campinas, CEP 13083-872, Campinas, São Paulo, Brazil}
\author{Paloma V. Silva}
\affiliation{Departamento de F\'isica, Universidade Federal do Piau\'i, CEP 64049-550, Teresina, Piau\'i, Brasil}
\author{Eduardo C. Gir\~ao}
\affiliation{Departamento de F\'isica, Universidade Federal do Piau\'i, CEP 64049-550, Teresina, Piau\'i, Brasil}
\author{Alexandre C. Dias}
\affiliation{Institute of Physics, University of Bras{\'{i}}lia, Bras{\'{i}}lia, 70910-900, Brazil}
\affiliation{International Center of Physics, University of Bras{\'{i}}lia, Bras{\'{i}}lia, 70910-900, Brazil}
\author{Douglas S. Galv\~ao}
\affiliation{Department of Applied Physics and Center for Computational Engineering and Sciences, State University of Campinas, CEP 13083-872, Campinas, São Paulo, Brazil}
\affiliation{Computational Materials Laboratory, LCCMat, Institute of Physics, University of Bras{\'{i}}lia, $70910-900$, Bras{\'{i}}lia, Brazil}
\author{Luiz A. R. Junior}
\affiliation{Institute of Physics, University of Bras{\'{i}}lia, Bras{\'{i}}lia, 70910-900, Brazil}
\affiliation{Computational Materials Laboratory, LCCMat, Institute of Physics, University of Bras{\'{i}}lia, $70910-900$, Bras{\'{i}}lia, Brazil}
 \date{\today}
 
 \begin{abstract}
 \noindent
Recently, two-dimensional goldene, a single-atomic-layer gold sheet, has been experimentally realized and is expected to be a promising material for next-generation devices, rather than three-dimensional bulk gold. Because quasi-1D derivatives can exhibit properties markedly different from their 2D counterparts, we investigate the stability, electronic, and magnetic properties of goldene-based nanoribbons (GOLDNRs) using first-principles computational methods. Our findings demonstrate that the electronic structure of these nanoribbons depends strongly on both edge chirality and ribbon width. Depending on the boundary configuration, the system transitions between semiconducting and metallic regimes, with electronic band gaps of $\approx 0.77$~eV and $\approx 0.04$~eV, or exhibits pristine metallic behavior. Additionally, the edge regions serve as highly electronically active sites, which may significantly enhance chemical reactivity relative to the central region. Furthermore, as the ribbon width increases, a clear transition occurs in which the metallic characteristics of the infinite sheet are recovered as the influence of quantum confinement and edge effects on the central domain is progressively mitigated. These results provide critical insights into tuning the physical properties of goldene via lateral confinement, highlighting its strong potential for applications in nanoscale field-effect transistors, spintronic devices, and catalysis.
\end{abstract}

\maketitle

\section{Introduction}
The discovery of two-dimensional (2D) materials has revolutionized condensed matter physics, enabling the exploration of exotic structural, electronic, optical, and mechanical properties at the atomic scale\cite{Kashiwaya2024, Mortazavi2024}. Among these systems, graphene remains the benchmark, demonstrating how reduced dimensionality and quantum confinement can drastically alter material behavior\cite{Joudi2015}. Motivated by the success of graphene, extensive research has shifted toward methods to synthesize other two-dimensional(2D) nanosystems, including other 2D carbon-based structures\cite{10.1021/jacs.2c06583,doi:10.1073/pnas.2413194122,10.1039/d5ta01619c}, hexagonal boron nitride\cite{kubota2007deep,song2010large}, graphitic carbon nitride\cite{thomas2008graphitic,algara2014triazine, 10.1039/c6ra14273g}, silicene\cite{lin2012structure,lin2013substrate, 10.1039/c4cp02902j}, germanene\cite{bianco2013stability}, stanene\cite{zhu2015epitaxial}, and phosphorene\cite{liu2014phosphorene,das2014ambipolar}, as well as transition metal dichalcogenides like $\text{MoS}_2$, $\text{WS}_2$, and $\text{ReS}_2$\cite{geim2013van,wang2012electronics, ReS2natue}. Despite these advances, synthesizing stable 2D materials composed solely of metallic elements remains a significant challenge. This paradigm has motivated the search for purely metallic 2D analogs, aiming to uncover novel physics and potential device applications. 

A remarkable recent breakthrough in this field was the experimental realization of \textit{goldene}, a strictly single-atom-thick gold monolayer~\cite {Kashiwaya2024}. Although Sharma et al.~\cite{Sharma2022} previously claimed goldene synthesis via thermal dewetting on sapphire, high-resolution scanning transmission electron microscopy (HR-STEM) and the presence of Au islands strongly suggested multilayer formation. Before 2024, the state of the art for unconfined 2D gold was limited to two-atom-thick nanosheets synthesized via a methyl orange-assisted wet-chemical route~\cite{Ye2019}. Kashiwaya et al. overcame this scalability barrier by selectively wet-etching the nanolaminate precursor Ti$_3$AuC$_2$ and using surfactant-assisted exfoliation to stabilize the structure against curling and agglomeration~\cite{Kashiwaya2024}. The resulting goldene exhibits a $9\%$ lattice contraction compared to bulk gold and a distinct shift in the Au 4f binding energy, highlighting its unique electronic environment~\cite{Kashiwaya2024}. Complementary theoretical and molecular dynamics studies have also confirmed the intrinsic stability of this 2D phase and predicted remarkable anisotropic thermal and mechanical responses~\cite{10.1039/d4mh01319k, Mortazavi2024, Iyakutti2025}. These advances firmly establish goldene as a pioneer in the expanding family of 2D metals, with exotic properties fundamentally distinct from those of their bulk counterparts. Other stable single-layer metallic structures have also been recently proposed~\cite{https://doi.org/10.1016/j.cplett.2026.142676,10.1021/acsomega.5c01823,10.1039/d6ta00772d}

Besides goldene, other gold-based nanostructures worthy mentioning are quasi-one-dimensional (quasi-1D) derivatives, including gold nanowires (NWs)~\cite{nano15060428}, gold nanotubes (GNTs)~\cite{ono2024breakdowncontinuumelasticityelectronic}, and especially gold nanoribbons (GNRs)~\cite{Fan2015}. Characterized as freestanding atomic membranes or elongated ribbons within 2D frameworks~\cite{Zhao2020, Fan2015}, GNRs offer an ideal platform for exploring novel physics. \textit{In situ} transmission electron microscopy has shown that electron irradiation can drive single-atom-thick gold membranes to evolve into nanoribbon-like structures, offering a direct view into their stability~\cite{Zhao2020}. Additionally, stabilizing GNRs in unconventional phases, such as the $4\text{H}$ hexagonal polymorphs, highlights how low-dimensional confinement can preserve metastable arrangements~\cite{Fan2015}. Consequently, these systems are uniquely suited for investigating quantum confinement, edge-state physics, and phase stability in reduced-dimensionality metals.  

Despite these promising developments, a comprehensive GNR investigation is still lacking. In particular, their structural stability and electronic properties remain unexplored relative to their monolayer and bulk gold counterparts. The ability to tune their width, chirality, and edge terminations offers a broad platform for both fundamental physics and potential applications in nanoscale electronics, plasmonics, and catalysis~\cite{Mortazavi2024, PatinoGuerrero2023}.

Motivated by the success of previous studies on low-dimensional systems, including one-dimensional structures such as nanoribbons, we used first-principles simulations within the density functional theory (DFT) framework to systematically investigate GNR structural stability and their electronic and mechanical properties. We aim to assess their viability in nanoelectronics by demonstrating the tunability of their physical properties via one-dimensional quantum confinement, width scaling, and edge configurations. The remainder of this paper is organized as follows: the next section describes the computational methods used, followed by a discussion of the structural geometry of the systems studied. Next, we present and analyze the results; finally, we summarize our findings and conclusions.

\section{Methodology}

We performed electronic structure simulations using the well-known DFT-based SIESTA code~\cite{soler2002siesta}. We used the Perdew-Burke-Ernzerhof (PBE) exchange-correlation functional within the generalized gradient approximation (GGA) to treat electron exchange and correlation~\cite{perdew1996generalized}, obtain optimized structures, and expand the valence wavefunctions with a double-$\zeta$ plus polarization (DZP) basis set. We modeled the ionic potential using Optimized Norm-Conserving Vanderbilt (ONCV) pseudopotentials~\cite{hamann2013optimized}. We performed Brillouin zone integrations with 40 x 1 x 1 (1 x 24 x 1) k-point sampling symmetry and nonequivalent k-points as part of a Monkhorst-Pack grid~\cite{monkhorst1976special} for the studied systems with small (large) unit cells.

All the studied systems were separated
by a vacuum distance of 40~\AA~ and 20~\AA~, laterally and vertically, respectively, from their periodic images to avoid spurious interactions among mirror cells.
We used a mesh cutoff of 400 Ry to represent charge density in real space, and set the density-matrix convergence criterion to at least $10^{-4}$. We used the conjugate gradient method to optimize the structural relaxations until the maximum force on each atom was less than $10^{-2}$~eV/~\AA~ and the maximum stress tensor component was less than $0.1$~GPa. 

\subsection{STUDIED STRUCTURES}

\begin{figure}[t]
  \begin{center}
  \includegraphics[width=\columnwidth]{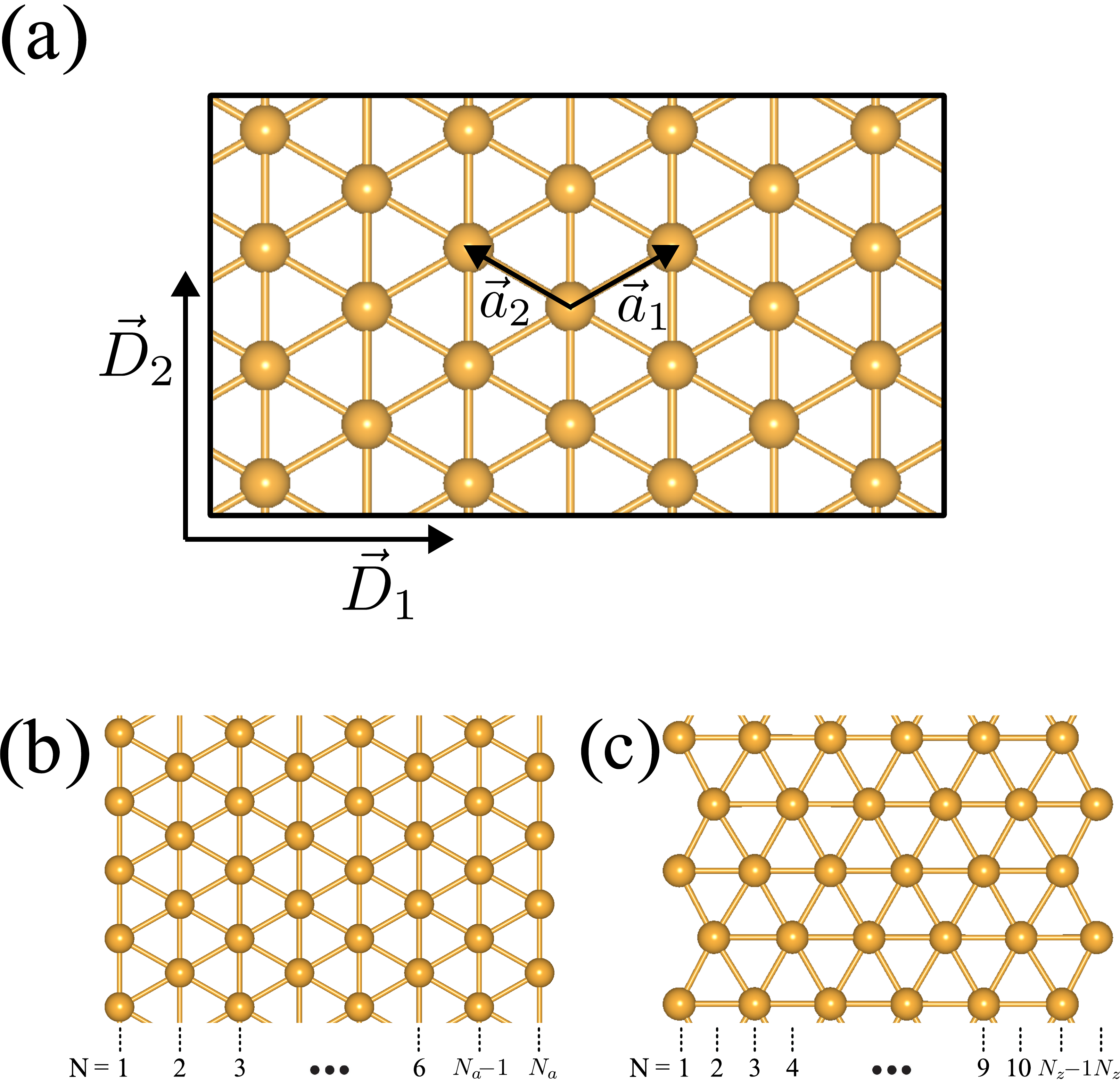}
  \caption{\footnotesize (a) Structural model of goldene, with $\vec{D}_{1}$ and $\vec{D}_{2}$ denoting the armchair and zigzag cutting orientations, respectively, and $\vec{a}_{1}$ and $\vec{a}_{2}$ the lattice vectors of the 2D goldene nanostructure. The families of AGOLDNRs considered here, with different edge configurations, are denoted as (b) a schematic of an 8-AGOLDNR and (c) a schematic of a 12-ZGOLDNR. The 1D unit-cell ribbon width is denoted by $N$, where $N_{a}$ ($N_{z}$) is the number of atoms along the armchair (zigzag) direction of the non-periodic ribbon.}
 \label{structures}
  \end{center}
 \end{figure}

Fig.~\ref{structures}(a) shows the top view of the crystalline structure of a goldene sheet, with the optimized lattice parameters $|\vec{a}_{1}| = |\vec{a}_{2}| = a = 2.59381$~\AA~ and an interaxial angle of $119.955^\circ$. These structural parameters agree well with previous experimental and theoretical reports~\cite{Yang2021, Kashiwaya2024}. The unit cell contains a single Au atom bonded to six nearest neighbors, forming a planar, honeycomb-like arrangement in a hexagonal lattice. The nanoribbons considered in this work can be conceptually derived by cutting the goldene sheet along the $\vec{D}_{1}$ and $\vec{D}_{2}$ crystallographic directions, defining the corresponding edge orientations. Cutting goldene along the $\vec{D}_{1} = \vec{a}_{1} - \vec{a}_{2}$ direction yields edges that we refer to as zigzag. This terminology follows from the orientation relative to the hexagonal rings and, by analogy with graphene, from the fact that the zigzag direction is also defined along $\vec{a}_{1} - \vec{a}_{2}$. Conversely, cutting along the $\vec{D}_{2} = \vec{a}_{1} + \vec{a}_{2}$ direction yields edges we designate as armchair, following the same reasoning used for zigzag nanoribbons.

Fig.~\ref{structures}(b) illustrates the schematic of an armchair goldene nanoribbon (AGOLDNR) with a width corresponding to $N_{a} = 8$ lines of dimer gold atoms along the non-periodic direction, denoted as an 8-AGOLDNR. Fig.~\ref{structures}(c) shows the schematic of a zigzag goldene nanoribbon (ZGOLDNR) with $N_{z} = 12$ lines of gold atoms along the non-periodic direction, denoted as a 12-ZGOLDNR. In general, the width of a goldene nanoribbon (GOLDNR) is defined by the number $N$ of gold atoms across its non-periodic direction. We have investigated systems with $N$ ranging from 2 to 12 for the armchair case and from 3 to 20 for the zigzag case. From this point onward, we will refer to a goldene nanoribbon containing $N_{a}$ atomic lines as an $N_{a}$-AGOLDNR, corresponding to the armchair case, and one containing $N_{z}$ atomic lines as an $N_{z}$-ZGOLDNR, corresponding to the zigzag case.

 Each gold atom in the goldene sheet is bonded to six nearest neighbors. However, in nanoribbons, edge atoms have unfulfilled dangling bonds. In the armchair configuration, all edge atoms have two dangling bonds, whereas in the zigzag configuration, the number of dangling bonds alternates between one and three. This distinctive bonding arrangement gives rise to additional electronic states localized at the nanoribbon edges, which we discuss in more detail in the following sections.

\section{Electronic properties}

\begin{figure}[h!]
  \begin{center}
  \includegraphics[width=\columnwidth]{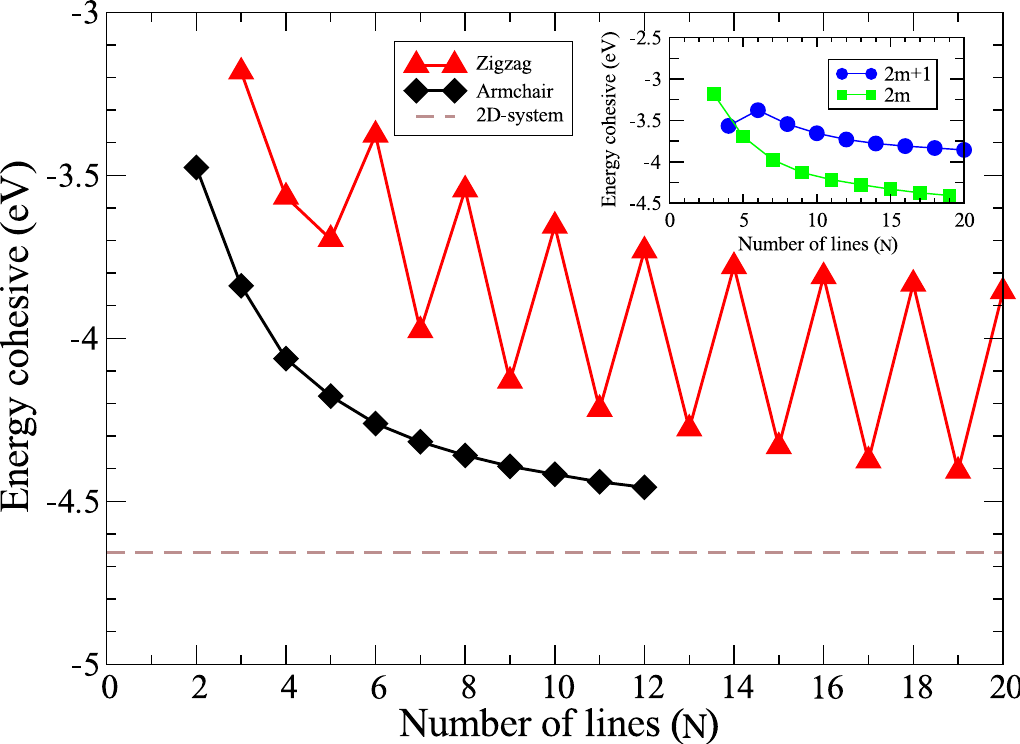}
  \caption{\footnotesize Estimated cohesive energy $E_{\mathrm{coh}}$ values as a function of width $N_{a}$ and $N_{z}$ for GOLDNR systems, namely ZGOLDNR (red line with triangles) and AGOLDNR (black line with squares). The $E_{\mathrm{coh}}$ value for the goldene sheet is represented by the brown dashed line at $E = -4.656$~eV. We calculate these values using Eq.~\ref{eq.coh}.}
 \label{Fig-Coh}
  \end{center}
 \end{figure}

We start our discussion with the estimated cohesive energy values of the proposed GOLDNRs and of their corresponding parent goldene sheet. Previous studies have shown that the parent goldene structure from which the GOLDNRs are derived is thermodynamically and dynamically stable, as confirmed by phonon dispersion and molecular dynamics simulations~\cite{sheremetyeva2025hydrogen}. To gain further insight into the structural stability of the GOLDNRs, we have computed their cohesive energies ($E_{\mathrm{coh}}$) according to the following equation:

\begin{equation}
    E_{coh}=\frac{E_{total}-N_{Au}\cdot E_{Au}}{N_{Au}},
    \label{eq.coh}
\end{equation}

and compared them with that of the goldene sheet. Here, $E_{\mathrm{total}}$ denotes the total energy of the system, $N_{\mathrm{Au}}$ is the total number of gold atoms in the unit cell, and $E_{\mathrm{Au}}$ represents the energy of an isolated gold atom. Fig.~\ref {Fig-Coh} shows the cohesive energies for the studied GOLDNRs and the goldene sheet. Our results suggest that the $N_{a}$-AGOLDNRs and $N_{z}$-ZGOLDNRs are energetically stable, with all cohesive energy values being negative and approaching that of the goldene sheet as the ribbon widths ($N_{a}$ and $N_{z}$) increase. For ZGOLDNRs (see inset of Fig.~\ref{Fig-Coh}), the $E_{\mathrm{coh}}$ approaches the 2D limit in an oscillatory manner, which can be classified into two distinct families: $N_{z} = 2m+1$ and $N_{z} = 2m$ (where $m$ is a positive integer). In contrast, for AGOLDNRs, this approach is monotonic.

\begin{figure}[h]
  \begin{center}
  \includegraphics[width=\columnwidth]{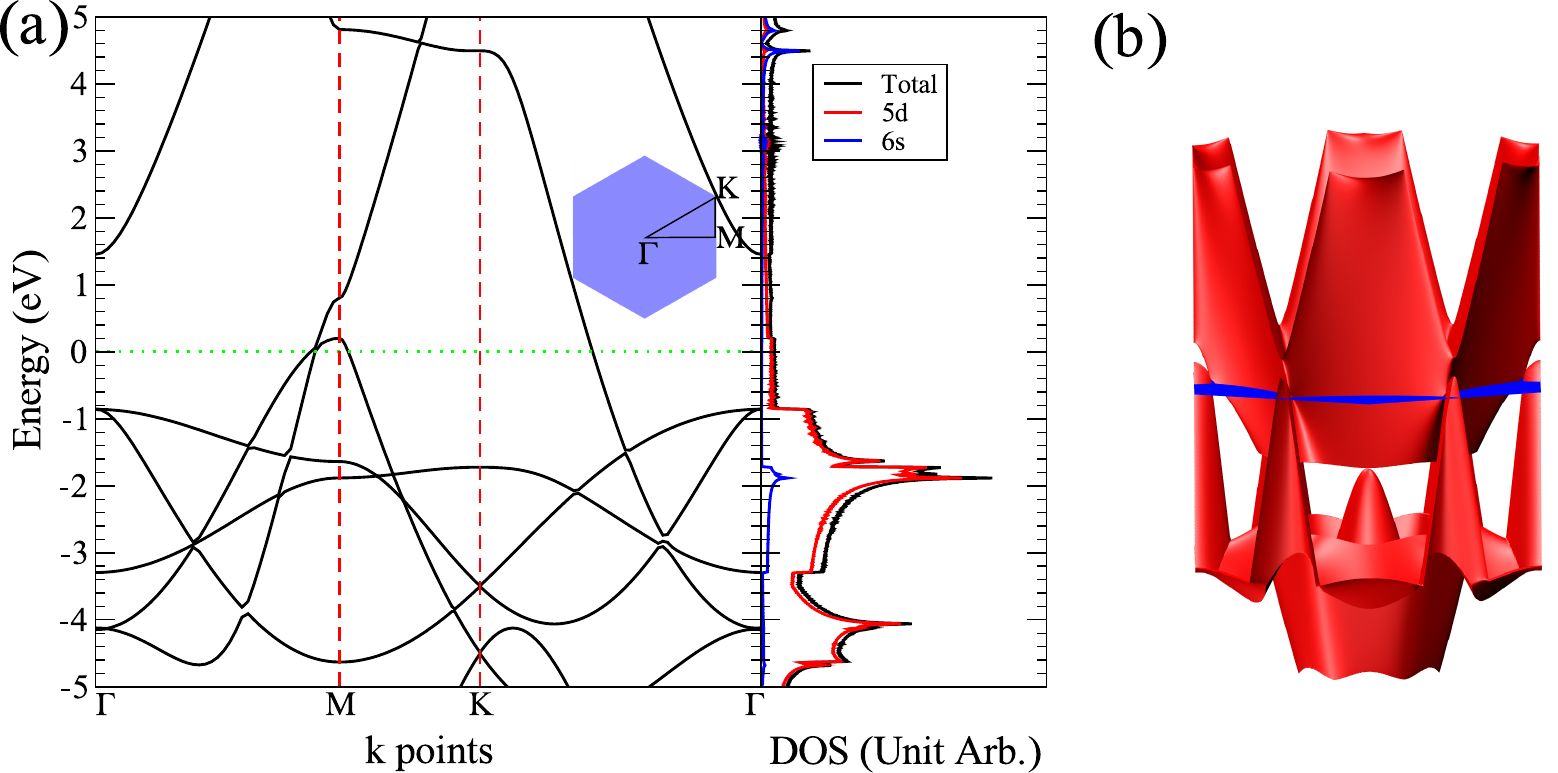}
  \caption{\footnotesize Electronic band structure of the goldene monolayer system at selected high-symmetry $k$ points, and the corresponding density of states (DOS) and projected density of states (PDOS) for the $5d$ and $6s$ orbitals as a function of energy. The Fermi level ($E_F$) was translated to the origin, and (b) the valence and conduction bands (red surfaces) are shown over the entire BZ.}
 \label{Fig-01}
  \end{center}
 \end{figure}

We now discuss the electronic structure of the goldene sheet. Figure~\ref{Fig-01}(a) displays its electronic band structure along the high-symmetry directions of the Brillouin zone (BZ), together with the corresponding projected density of states (PDOS). The valence and conduction bands cross near the $M$ point and midway along the $\Gamma$–$K$ path, confirming the monolayer's metallic character, in close analogy with the $\chi_3$-borophene sheet~\cite{vishkayi2018freestanding}. As in $\beta_{12}$-borophene~\cite{feng2017dirac} and monolayer $S_bSS_n$~\cite{liu2025study}, several Dirac-like cones appear well below the Fermi level in the valence region. Previous theoretical studies have shown that tensile strain can shift Dirac-like crossings toward the Fermi level in two-dimensional materials such as monolayer $S_bSS_n$~\cite{liu2025study}, suggesting that similar effects may also occur in goldene under suitable external perturbations. The PDOS shows that the electronic states near the Fermi level are predominantly derived from Au $5d$ orbitals, consistent with recent theoretical studies on goldene~\cite{dos2025exploring,li2025structural}. As shown in Fig.~\ref{Fig-01}(b), the three-dimensional electronic band structure over the entire Brillouin zone highlights multiple band crossings at the zone center and corners, further confirming the metallic electronic structure of goldene and providing the basis for understanding the electronic properties of the corresponding nanoribbons discussed in the following sections.

We will now discuss the armchair goldene nanoribbons ($N_{a}$-AGOLDNR) results. As will be shown, all nanoribbons defined along this crystallographic direction exhibit metallic behavior. This characteristic is directly linked to the electronic properties of the 2D system. In particular, the reciprocal space of an AGOLDNR corresponds to the $\Gamma$–$K$ direction of the 2D Brillouin zone, along which the goldene sheet bands cross the $E_F$ at the midpoint of the path~\cite{sheremetyeva2025hydrogen}. In this sense, the metallic behavior observed in AGOLDNRs (as will be discussed below) can be directly associated with the electronic properties of the 2D system. This occurs mainly because edge contributions and boundary conditions do not solely dominate the low-energy states. In this system, all edge atoms of armchair ribbons have a uniform coordination number ($CN=4$), making the edge neighborhood nearly identical to that at the center. Nevertheless, both edge and quantum confinement effects associated with the ribbon width play a nontrivial role in determining the overall electronic structure.

\begin{figure}[h!]
  \begin{center}
  \includegraphics[width=\textwidth]{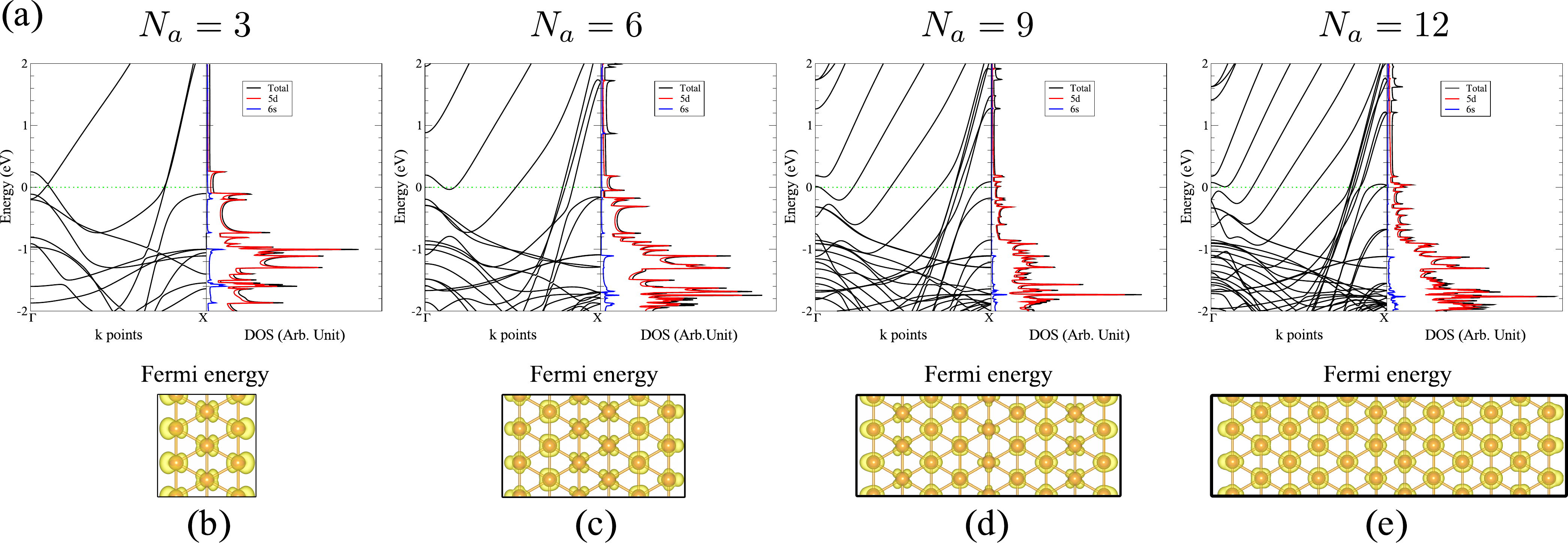}
  \caption{\footnotesize (a) Electronic band structure with the corresponding density of states (DOS) and projected density of states (PDOS) for the $5d$ and $6s$ orbitals of 3-AGOLDNRs, 6-AGOLDNRs, 9-AGOLDNRs, and 12-AGOLDNRs (Fermi level ($E_F$) is at $E = 0$). The local density of states (LDOS) at the Fermi energy ($E_{F}$) for $N_{a}$-AGOLDNRs with (b) $N_{a}=3$, (c) $N_{a}=6$, (d) $N_{a}=9$, and (e) $N_{a}=12$, respectively.}
 \label{Fig-03}
  \end{center}
 \end{figure}

Figure~\ref{Fig-03}(a) shows the electronic band structures of the $N_{a}$-AGOLDNRs with $N_a = 3, 6, 9, \text{ and } 12$, respectively. The green dotted line indicates $E_{F}$ at $E=0$. Near the $\Gamma$ point of the BZ, the $N_{a}=3$ ribbon displays a cone-like band intersection. Additionally, near the opposite edge of the BZ boundary, at the $X$ point, two bands cross the $E_{F}$. The PDOS shows that these frontier states mainly derive from the $5d$ orbitals. As the ribbon width increases, more bands cross $E_{F}$. This indicates that we can tune the number of available channels for electronic transport by changing the nanoribbon width. The frontier states near $E_F$ are spatially distributed across the ribbon width. Also, the systems have a pronounced contribution from highly localized edge states [see Figs.~\ref{Fig-03}(b)-(e)].

 \onecolumngrid
 
 \twocolumngrid

\begin{figure}[h!]
  \begin{center}
\includegraphics[width=0.9\textwidth]{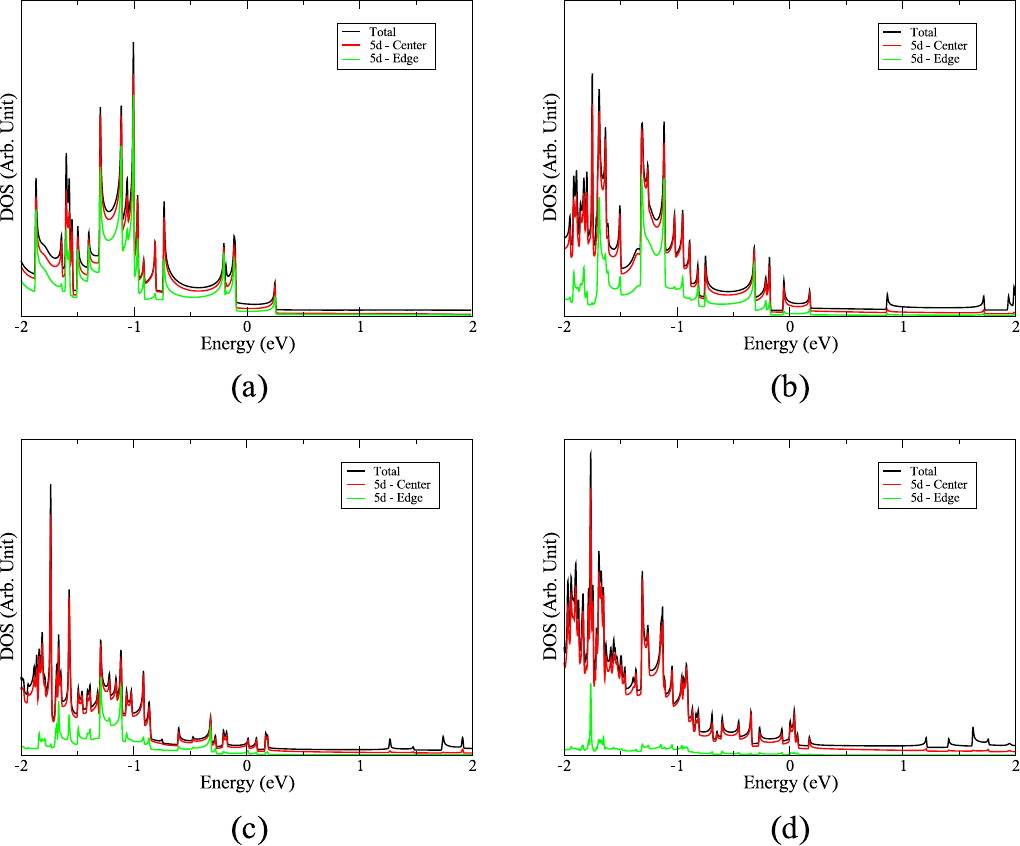}
\caption{\footnotesize Total density of states (DOS) and projected density of states (PDOS) for the $N_{a}$-AGOLDNR systems with (a) $N_{a}=3$, (b) $N_{a}=6$, (c) $N_{a}=9$, and (d) $N_{a}=12$. The curves represent the total DOS (black line) and the PDOS contributions from the edge (green line) and central (red line) gold atoms to the $5d$ orbital.}
\label{Fig-03-1}
  \end{center}
 \end{figure}

To further investigate how edge effects influence the electronic properties of AGOLDNRs, Fig.~\ref{Fig-03-1} shows the PDOS of the $5d$ orbitals projected onto the central and edge atoms, focusing on the energy range around $E_F$. As observed in Figs.~\ref{Fig-03-1}(a)–(d), for narrower widths, the $5d$ orbitals near $E_F$ contribute in a manner highly similar to the PDOS profiles of the edge and central regions. However, as the ribbon width increases, the central atoms' PDOS gradually evolves toward bulk-like characteristics. This evolution, in contrast to the persistent and localized edge states, provides compelling evidence that quantum confinement effects are progressively mitigated as the system approaches the 2D monolayer limit. For narrow widths, the conduction band region is characterized by a few distributions of discrete, well-defined peaks. As the ribbon width increases, the number of peaks also increases. This increase in the distribution of the peaks within the conduction region, coupled with the persistent metallic behavior at the Fermi level, demonstrates that while the system remains metallic, its electronic structure is highly sensitive to lateral scaling, which dictates the discretization of the unoccupied states. The observed trends are consistent across all AGOLDNR widths studied, with the evolution from a few unoccupied states to many unoccupied states providing clear evidence of the transition from a strongly confined regime to a bulk-like 2D continuum (see Fig.~\ref{fig:si_fig1}).

 \onecolumngrid
 
 \twocolumngrid

 We also investigated the electronic distribution between $5d$ orbitals at the edges and at the center using Mulliken population analysis of the AGOLDNRs. In this analysis, we selected the outermost atom to represent the edge (E) site, while the atom immediately adjacent to it, oriented toward the center of the nanoribbon, was defined as the central (C) site. The results, summarized in Table~\ref{tab-milli}, reveal individual orbital occupancies that highlight pronounced anisotropy in the $5d$ electron states. We observe that the $d_{xy}$ and $d_{x^2-y^2}$ orbitals show the largest variations in occupancy as we move from the center to the edge. These orbitals are distributed in the ribbon plane, and confinement forces electrons to jump between specific orbital channels to accommodate the reduced lateral space. This difference reflects strong boundary-induced ligand-field effects, where the reduced atomic environment at the edge favors the occupation of specific orbitals. When considering the integrated contribution, we confirm that edge sites consistently accumulate a higher total $5d$ electron density ($\approx 9.25$~$e$) than the center ($\approx 9.16$~$e$). 
 
 This charge accumulation is an electronic response to reduced atomic coordination at nanoribbon boundaries, which requires redistributing the electron cloud to minimize the system’s total energy. Furthermore, our results demonstrate fast electronic convergence. The central population ($\sum Q_{5d, Center}$) stabilizes at $\approx 9.164$~$e$ starting from width $N_{a}=6$, indicating that central atoms are effectively decoupled from boundary effects and exhibit bulk-like electronic behavior. In this system, the constant charge difference ($Q_{5d, Center}-Q_{5d, Edge} \approx 0.09$~$e$) indicates that AGOLDNRs exhibit a unique electronic profile, making the boundaries chemically distinct and more reactive  (as expected) than the center. These findings suggest that the edge regions serve as electronically active sites that may significantly enhance chemical reactivity relative to the central region.
 
 \onecolumngrid
 
 \begin{table}[h!]
\centering
\renewcommand{\arraystretch}{1.2}
\begin{tabular}{lccccccccccc}
\hline
\textbf{System} & \multicolumn{2}{c}{\textbf{$d_{xy}$}} & \multicolumn{2}{c}{\textbf{$d_{yz}$}} & \multicolumn{2}{c}{\textbf{$d_{z^2}$}} & \multicolumn{2}{c}{\textbf{$d_{xz}$}} & \multicolumn{2}{c}{\textbf{$d_{x^2-y^2}$}} & \textbf{$\sum Q_{5d}$} \\ \cline{2-11}
 & E & C & E & C & E & C & E 
 & C & E & C & (E / C) \\ \hline
$N_{a}=3$   & 1.694 & 1.775 & 1.972 & 1.960 & 1.884 & 1.858 & 1.945 & 1.946 & 1.735 & 1.636 & 9.230 / 9.175 \\
$N_{a}=6$   & 1.764 & 1.728 & 1.968 & 1.954 & 1.881 & 1.862 & 1.948 & 1.949 & 1.696 & 1.671 & 9.257 / 9.164 \\
$N_{a}=9$   & 1.771 & 1.734 & 1.969 & 1.953 & 1.882 & 1.862 & 1.949 & 1.950 & 1.685 & 1.665 & 9.256 / 9.164 \\
$N_{a}=12$ & 1.767 & 1.732 & 1.968 & 1.953 & 1.881 & 1.863 & 1.950 & 1.950 & 1.692 & 1.665 & 9.258 / 9.163 \\ \hline
\end{tabular}
\caption{Evolution of effective Mulliken populations for $N_a$-AGOLDNRs of increasing width ($N_{a}=3$ to $N_{a}=12$) to the $5d$ orbital. E = Edge, C = Center.}
\label{tab-milli}
\end{table}

\twocolumngrid

\begin{figure}[h!]
  \begin{center}
  \includegraphics[width=\textwidth]{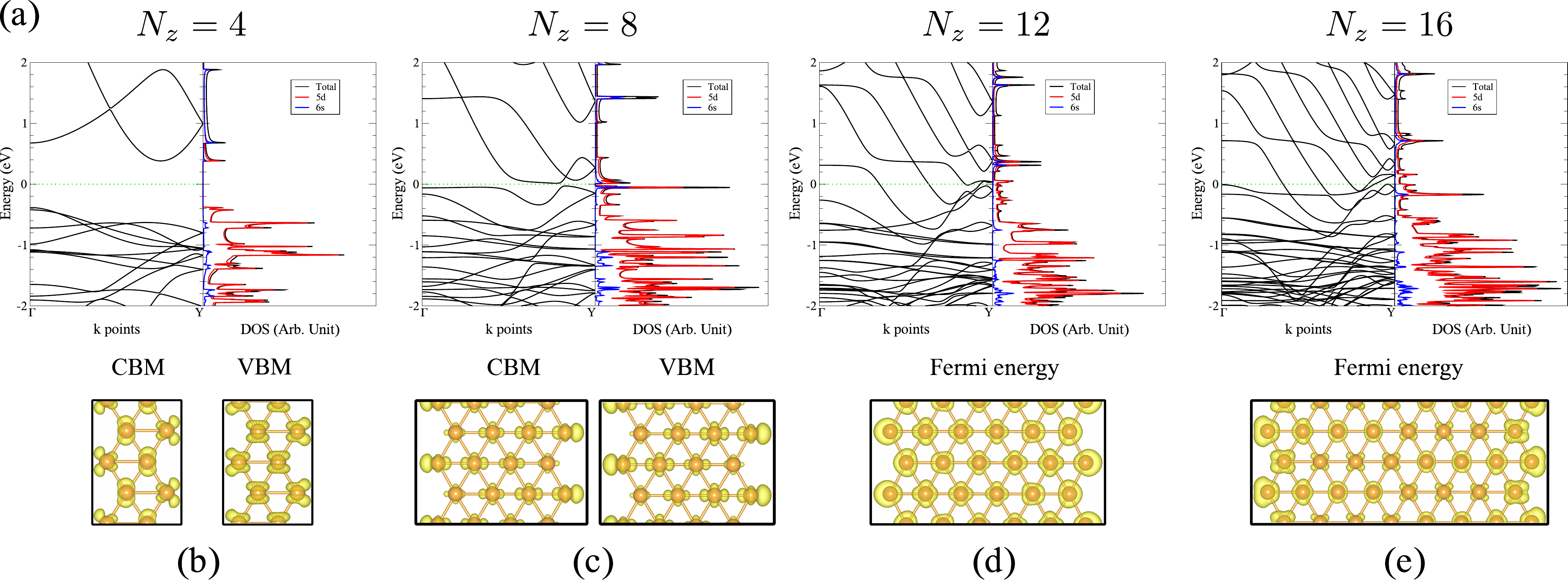}
  \caption{\footnotesize (a) Electronic band structure with density of states (DOS) and projected density of states (PDOS) for the $5d$ and $6s$ orbitals of 4-ZGOLDNRs, 8-ZGOLDNRs, 12-ZGOLDNRs, and 16-ZGOLDNRs ($E_F$ is at $E = 0$). The local density of states (LDOS) at $E_F$ energy for $N_{a}$-ZGOLDNRs with (b) $N_{z}=4$, (c) $N_{z}=8$, (d) $N_{z}=12$, and (e) $N_{z}=16$, respectively. }
 \label{Fig-04}
  \end{center}
 \end{figure}

We now extend our analysis to the zigzag-edged nanoribbon (ZGOLDNR) family. Fig.~\ref{Fig-04} shows the electronic band structures, projected density of states (PDOS), and local density of states (LDOS) for representative systems ($N_z = 4, 8, 12, \text{ and } 16$), illustrating how electronic properties evolve with ribbon width. The electronic band structures reveal distinct features that contrast significantly with those of armchair counterparts, particularly in state localization and energy distribution near the $E_F$ in narrow ribbons. As shown, the reciprocal space of ZGOLDNRs corresponds to the $\Gamma$–$M$ direction of the 2D BZ, along which the goldene sheet bands cross $E_F$ near the $M$ point~\cite{sheremetyeva2025hydrogen}. Furthermore, because not all allowed transverse wave vectors intersect the electronic states at $E_F$, this geometry exhibits remarkably rich electronic behavior.

\begin{figure}[h!]
  \begin{center}
  \includegraphics[width=0.7\columnwidth]{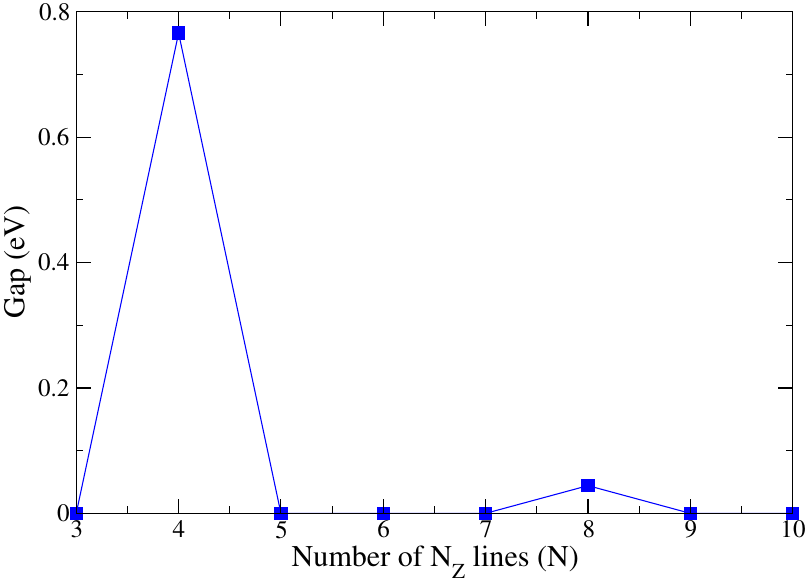}
  \caption{\footnotesize Electronic band gap of semiconducting ZGOLDRNs as a function of their width $N$.}
 \label{gap}
  \end{center}
 \end{figure}

The electronic behavior of goldene nanoribbons is intrinsically governed by the boundary coordination environment. While armchair ribbons feature a uniform coordination number ($CN=4$) for all edge atoms, zigzag ribbons exhibit an alternating structural motif. In the latter case, the outermost atoms have a low coordination number ($CN=3$), while the adjacent inner atoms have $CN=5$.  This structural asymmetry fundamentally dictates the electronic properties. As shown in Fig.~\ref{Fig-04}(a), the $N_z=4$ system exhibits an indirect electronic band gap of $\approx 0.77$~eV. This result highlights the competition between quantum confinement and boundary coordination. In this regime, the latter opens a semiconducting gap, in contrast to the $N_z=3$ configuration, where quantum confinement predominates, yielding metallic behavior analogous to narrow AGOLDNR systems. Although not the narrowest configuration, this system highlights the critical competition between quantum confinement and the boundary coordination environment. For larger widths, the band gap values oscillate, alternating between semiconducting and metallic states, as seen at $N_z=8$, where the indirect gap is $\approx 0.04$~eV, as illustrated in Fig.~\ref{gap}.

The interplay between edge-state density and quantum confinement can explain this phenomenon. These oscillations arise from the inherent modulation of the electronic symmetry at the zigzag edges, where the parity of the nanoribbon width governs the overlap of localized edge states. As the ribbon width increases to $N_z = 12$, the system recovers the metallic characteristics of the 2D bulk. However, it shows more states crossing $E_F$, significantly increasing the number of available channels for electronic transport (see Fig.~\ref{fig:si_fig2}). As in the armchair case, these frontier states derive primarily from the $5d$ orbitals, as confirmed by the PDOS. However, in contrast to the armchair geometry, these states are primarily localized at the edges in narrower ribbons. The DOS peaks, coinciding with the quasi-flat bands near the $E_F$, provide clear evidence of edge-induced states that dominate the low-energy electronic properties in narrow systems. As the width increases, these states rapidly redistribute throughout the ribbon, reflecting the ready convergence toward 2D system-like electronic properties. These results are corroborated by LDOS for levels near $E_F$, as shown in Figs.~\ref{Fig-04}(b)–(e), which demonstrate that frontier states transition from edge-confined to a more delocalized distribution across the nanoribbon. These findings show that goldene at low dimensionality offers a robust platform for precisely tuning electronic properties through rigorous control of the geometric edge and nanoribbon width ($N_{z/a}$).

 \onecolumngrid

\twocolumngrid

\begin{figure}[h!]
  \begin{center}
  \includegraphics[width=0.9\textwidth]{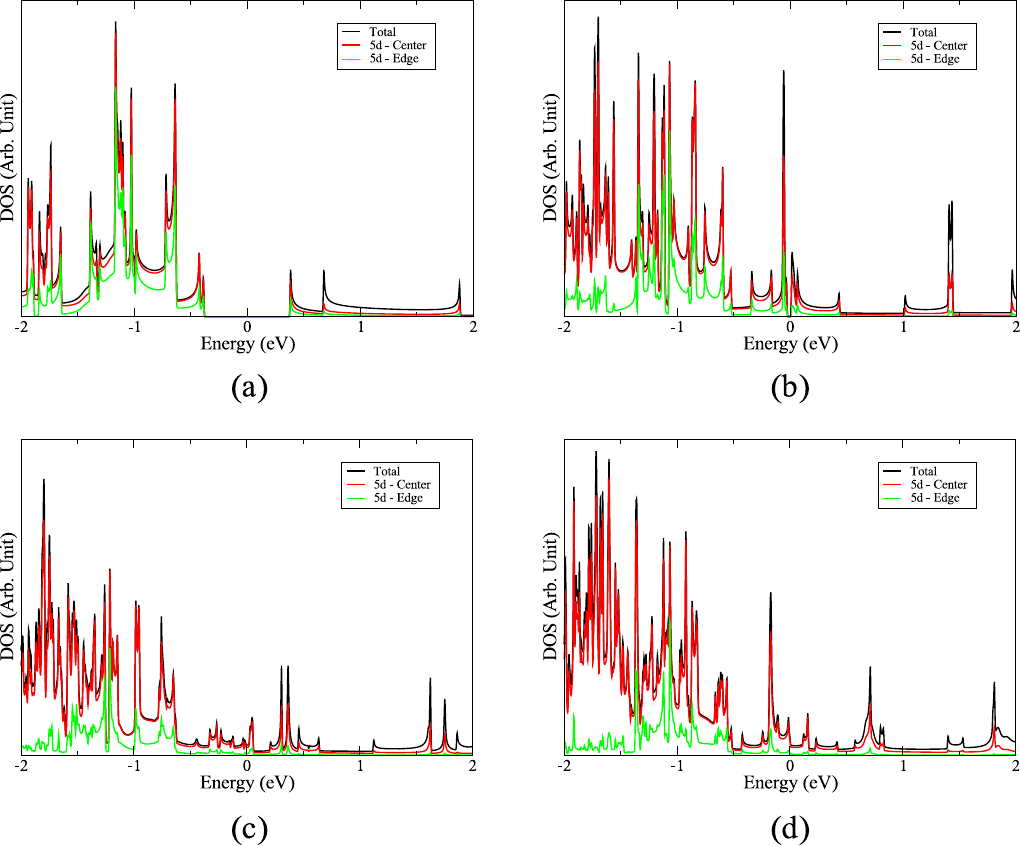}
  \caption{\footnotesize Total density of states (DOS) and projected density of states (PDOS) for the $N_{z}$-ZGOLDNR systems with (a) $N_{z}=4$, (b) $N_{z}=8$, (c) $N_{z}=12$, and (d) $N_{z}=16$. The curves represent the total DOS (black line) and the PDOS contributions from the edge (green line) and central (red line) gold atoms to the $5d$ orbital.}
 \label{Fig-05}
  \end{center}
 \end{figure}

To investigate how edge and central regions influence the electronic properties of ZGOLDNRs near $E_F$, Fig.~\ref{Fig-05} shows the $5d$-projected density of states (PDOS) for central and edge atoms. As observed in Figs.~\ref{Fig-05}(a)-(d), for smaller widths, the contributions from both regions are remarkably similar near the $E_F$. However, as the ribbon width increases, the PDOS of the central atoms gradually evolves, reflecting a transition toward the electronic properties of the 2D system.

 \onecolumngrid

\twocolumngrid

 \onecolumngrid

 \twocolumngrid

Finally, we analyzed the electronic distribution between the edges and the center through the Mulliken population analysis in the ZGOLDNR geometry (see Table~\ref{tab:zigzag-mulliken-evol}). We categorized the structural sites based on their position relative to the ribbon width. The edge (E) site is the outermost atomic row along the zigzag termination, with a low coordination number of $CN=3$. The center (C) site is defined as the atomic row immediately adjacent to the edge, positioned toward the interior of the nanoribbon. This distinction allows for a clear comparison of electronic contributions between the highly strained boundary region and the ribbon's inner region. 

Unlike armchair geometry, which maintains a relatively uniform charge distribution, the zigzag configuration imposes a distinct asymmetric electronic environment. In the narrow ZGOLDNR ($N_{z}=4$), the low coordination ($CN=3$) at the boundaries induces a significant geometric symmetry breaking, forcing a massive accumulation of electronic charge at the edge sites ($\sum Q_{5d} = 9.358$ $e$ versus $9.178$ $e$ at the center). As detailed in Table~\ref{tab:zigzag-mulliken-evol}, this spatial redistribution is primarily accommodated by the in-plane $5d$ orbitals, where the $d_{xy}$ populations reach $1.715$ $e$ (E) and $1.606$ $e$ (C), while the $d_{x^2-y^2}$ populations are $1.836$ $e$ (E) and $1.797$ $e$ (C). This large localized charge accumulation generates a strong local electrostatic potential at the boundaries, drastically altering the hybridization of the frontier states near $E_F$. Consequently, the local degeneracy of the electronic bands is lifted, splitting the states into distinct valence and conduction bands, which is the fundamental mechanism for opening the electronic gap. 

As the nanoribbon width scales up to $N_{z}=12$ and $N_{z}=16$, the zigzag systems do not follow a simple charge-smoothing trend but instead maintain a robust, nontrivial asymmetry between the edge and the center. For the $N_{12}$ width, the $d_{xy}$ population at the edge actually increases to $1.823$ $e$ while the central region reaches $1.698$ $e$, yielding a maximum orbital anisotropy of $\Delta Q_{d_{xy}} = d_{xy}^{\text{E}} - d_{xy}^{\text{C}} \approx 0.125$~ $e$. This localized charge behavior at the zigzag boundaries is consistently present. 

Interestingly, the electronic band structures of the ZGOLDNR family show a pronounced band-braiding effect near $E_F$, which becomes clearer in wider systems ($N_z > 10$) (see Fig.~\ref{fig:si_fig2}). Rather than being restricted to extreme confinement, this phenomenon stabilizes as the ribbon width increases, highlighting its robustness in ZGOLDNRs. This band-braiding effect is consistent with the tight-binding model proposed by Correa et al.~\cite{Correa2018} for zigzag nanoribbons, which attributes such crossings to third-nearest-neighbor hopping ($t_3$). Furthermore, independent DFT studies on chevron-type ribbon derivatives~\cite{Metallic_Chevron_2020} corroborate the formation of these complex, symmetry-protected crossings. In ZGOLDNRs, these features persist in wider ribbons because the low edge coordination number ($CN=3$) accumulates electronic charge in the in-plane $5d_{xy}$ and $5d_{x^2-y^2}$ orbitals, driving continuous rehybridization of the frontier states.

As the width expands further, the ribbon's central region gradually recovers a pristine 2D crystalline environment, becoming effectively screened from edge-induced coordination effects. Our results indicate that the spatial delocalization of the electronic states governs this transition toward 2D metallic behavior. While the boundary states remain strictly localized at the $CN=3$ edges, the expanding central domain accommodates extended states, progressively mitigating quantum confinement and restoring the electronic features of the goldene sheet.

 \onecolumngrid
 
 \begin{table}[h!]
\centering
\renewcommand{\arraystretch}{1.2}
\begin{tabular}{lccccccccccc}
\hline
\textbf{System} & \multicolumn{2}{c}{\textbf{$d_{xy}$}} & \multicolumn{2}{c}{\textbf{$d_{yz}$}} & \multicolumn{2}{c}{\textbf{$d_{z^2}$}} & \multicolumn{2}{c}{\textbf{$d_{xz}$}} & \multicolumn{2}{c}{\textbf{$d_{x^2-y^2}$}} & \textbf{$\sum Q_{5d}$} \\ \cline{2-11}
 & E & C & E & C & E & C & E & C & E & C & (E / C) \\ \hline
$N_{z}=4$    & 1.715 & 1.606 & 1.960 & 1.944 & 1.877 & 1.865 & 1.970 & 1.966 & 1.836 & 1.797 & 9.358 / 9.178 \\
$N_{z}=8$    & 1.765 & 1.656 & 1.963 & 1.950 & 1.878 & 1.863 & 1.966 & 1.953 & 1.747 & 1.729 & 9.319 / 9.151 \\
$N_{z}=12$ & 1.823 & 1.698 & 1.961 & 1.955 & 1.869 & 1.865 & 1.965 & 1.951 & 1.712 & 1.661 & 9.330 / 9.130 \\
$N_{z}=16$ & 1.786 & 1.677 & 1.961 & 1.953 & 1.878 & 1.866 & 1.965 & 1.952 & 1.732 & 1.682 & 9.322 / 9.130 \\ \hline
\end{tabular}
\caption{Evolution of effective Mulliken populations for $N_z$-ZGOLDNRs of increasing width ($N_{z}=4$ to $N_{z}=16$) to the $5d$ orbital. E = Edge, C = Center.}
\label{tab:zigzag-mulliken-evol}
\end{table}

 \twocolumngrid

\section{Spin-Polarized States}

A noteworthy aspect of the electronic properties of both AGOLDNRs and ZGOLDNRs is the high charge density localized at the edges and the presence of flat bands near $E_F$ in specific systems. These features strongly favor spin-polarized (SP) configurations, in direct analogy to zigzag-edged graphene~\cite{son2006energy}, $\chi_3$-borophene~\cite{vishkayi2018freestanding}, and $\text{NbOCl}_2$ nanoribbons~\cite{ghasemi2026tailoring}. 

\begin{figure}[tbp]
    \centering

    \includegraphics[
        width=0.95\columnwidth
    ]{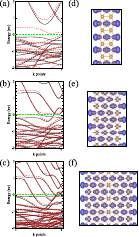}

    \caption{%
    Electronic band structure of
    (a) 5-ZGOLDNRs,
    (b) 8-ZGOLDNRs, and
    (c) 13-ZGOLDNRs in the
    ferromagnetic (FM) configuration,
    and spin-polarized charge distribution
    for the FM configuration of
    (d) 5-ZGOLDNRs,
    (e) 8-ZGOLDNRs, and
    (f) 13-ZGOLDNRs, respectively.
    Blue colors denote spin-up components.
    The isosurface value is
    $0.05\,e\,\text{\AA}^{-1}$.
    }

    \label{Spin}
\end{figure}

Although gold atoms in monolayer membranes are intrinsically diamagnetic, dimensionality reduction dramatically alters this picture. Wang et al. reported that previous first-principles studies found emergent ferromagnetism in ultra-narrow gold nanoribbons, yielding magnetic moments of $0.38\text{--}0.51\,\mu_B$ per unit cell for ribbon widths below $0.9\,\text{nm}$~\cite{wang2019free}.
By explicitly incorporating the spin degree of freedom in our DFT calculations, we find that only the ZGOLDNR family converges to a spin-polarized ground state, specifically a ferromagnetic (FM) configuration. As shown in Fig.~\ref{Spin}, the spins align ferromagnetically along each edge, with a parallel FM edge-to-edge coupling~\cite{pisani2007electronic,wang2019free}. In Fig.~\ref{Spin}, we show the spin-resolved band structures for the representative 5-, 8-, and 13-ZGOLDNRs, where spin-up and spin-down states are represented by full black and dashed red curves, respectively. All other width configurations converge to non-polarized (NP) systems and exhibit band structures similar to those of NP systems (see Fig.~\ref{fig:si_fig3}).

For the three systems exhibiting an FM ground state, the spin-up and spin-down bands are non-degenerate (see Figs.~\ref{Spin}(a)-(c)). To analyze this spin splitting, we plot the spatial spin density (defined as $\Delta\rho = \rho_{up} - \rho_{down}$) in Figs.~\ref{Spin}(d)-(f), where blue and red isosurfaces represent majority spin-up and spin-down densities, respectively. Because the spin-up and spin-down electronic clouds are not spatially symmetric, simple spatial symmetry operations cannot map one onto the other. Consequently, this spatial inequivalence changes the spin degeneracy, resulting in the observed spin-split band structure.

Remarkably, this SP configuration induces a transition to semiconducting behavior only in the 5-ZGOLDNR system, with magnetic moments of $\approx~1.0~\mu_B$, in contrast to its metallic NP counterpart. In this specific width, the spin-splitting and the quantum confinement under reduced boundary conditions, even though this is not the narrowest ribbon, cooperatively open an indirect electronic band gap of $\approx0.09$~eV for the spin-down channel, and a larger indirect band gap of $\approx0.73$~eV for the spin-up channel, characterizing a promising spin-filter system~ \cite{silva2018high,silva2021electronic}. This band-gap opening arises because localized edge bands are selectively populated with a single spin component, shifting the opposite-spin states into unoccupied states. Notably, this shift preserves the low-dispersion characteristic of the frontier bands, leaving their almost flat-band shape highly reminiscent of the original NP states around $E_F$.

Conversely, the wider systems do not exhibit semiconducting behavior despite their spin polarization. In the 8-ZGOLDNR, although spin splitting is present, with magnetic moments of $\approx~0.59~\mu_B$, the strong influence of the edge electrons, combined with the increased ribbon width, closes the electronic gap found in the NP configuration and renders the system metallic. Similarly, the 13-ZGOLDNR retains its metallic character, with magnetic moments of $\approx~0.82~\mu_B$, although the bands are clearly non-degenerate.

For other system widths, the NP and SP band profiles remain similar, suggesting that spin polarization plays a negligible role in those systems. To understand the energetic stability of these magnetic configurations and how the interplay between quantum confinement and orbital edge hybridization scales with the ribbon width, we calculate the magnetization energy ($E_{\mathrm{mag}}$) for all ZGOLDNR systems~\cite{de2019electronic,maia2026electronic}. This quantity is defined as:

\begin{equation}
E_{\mathrm{mag}} = E_{\mathrm{FM}} - E_{\mathrm{NP}},
\label{eq:emag}
\end{equation}

where $E_{\mathrm{FM}}$ and $E_{\mathrm{NP}}$ are the total energies per unit cell for the ferromagnetic and non-polarized configurations, respectively. Consequently, negative $E_{\mathrm{mag}}$ values indicate that the FM state is energetically favored over the NP state, with more negative values indicating greater relative stability of the FM configuration.

\begin{table}[ht]
\centering
\begin{tabular}{cccc}
\hline
Width ($N$) & $E_{\mathrm{mag}}$ (meV) & Width ($N$) & $E_{\mathrm{mag}}$ (meV) \\
\hline
$N_{z}=3$  & 0.000   & $N_{z}=12$ & -0.010 \\
$N_{z}=4$  & 0.000   & $N_{z}=13$ & -22.832 \\
$N_{z}=5$  & -33.969 & $N_{z}=14$ & -1.463 \\
$N_{z}=6$  & 0.112   & $N_{z}=15$ & 0.002 \\
$N_{z}=7$  & -0.002  & $N_{z}=16$ & 3.659 \\
$N_{z}=8$  & -5.530  & $N_{z}=17$ & -0.017 \\
$N_{z}=9$  & -0.001  & $N_{z}=18$ & 7.016 \\
$N_{z}=10$ & -0.025  & $N_{z}=19$ & -0.002 \\
$N_{z}=11$ & 0.004   & $N_{z}=20$ & -0.003 \\
\hline
\end{tabular}
\caption{Estimated magnetization energy, defined as $E_{\mathrm{mag}} = E_{\mathrm{FM}} - E_{\mathrm{NP}}$, for the $N_z$-ZGOLDNRs. Negative values indicate that the ferromagnetic state is energetically more stable than the nonmagnetic state, whereas positive values indicate that the nonmagnetic state is more stable.}
\label{tab3}
\end{table}

Based on the estimated values listed in Table~\ref{tab3}, several key physical insights emerge. First, the magnetization energy does not scale monotonically with ribbon width but instead exhibits oscillations, with prominent energy minima at specific widths. The ferromagnetic ground state is more stable for $N_z = 5$ ($E_{\mathrm{mag}} = -33.969$~meV) and $N_z = 13$ ($E_{\mathrm{mag}} = -22.832$~meV), followed by a moderately stable metallic FM configuration at $N_z = 8$ ($E_{\mathrm{mag}} = -5.530$~meV). For these cases, particularly $N_z = 5$ and $13$, the magnitude of $E_{\mathrm{mag}}$ is comparable to or exceeds the thermal energy at room temperature ($k_BT~\approx25.7$~meV), suggesting that their magnetic configuration could be experimentally robust against thermal fluctuations. 

Conversely, for the ultra-narrow system ($N_z = 3$ and $4$), $E_{\mathrm{mag}}$ is zero, indicating that the extreme quantum confinement and the strong hybridization between the opposing edges completely extinguish the localized edge magnetic moments. For most other widths, $E_{\mathrm{mag}}$ fluctuates around negligible values (within a few microelectronvolts), showing that the FM and NP configurations are degenerate and the system preserves its non-polarized ground state.

\section{Concluding remarks}

In summary, we investigated the electronic properties of nanoribbons derived from recently synthesized two-dimensional goldene, systematically evaluating how lateral confinement and edge configurations shape their fundamental physical properties. Our first-principles calculations demonstrate that the electronic profile of these goldene nanoribbons (GOLDNRs) is exceptionally sensitive to edge chirality and width scaling. Depending on the boundary geometry, the systems can be tailored to transition from distinct semiconducting phases, exhibiting well-defined electronic band gaps of  $\approx 0.77$~eV and  $\approx 0.04$~eV, to a robust metallic state. Furthermore, our orbital analysis reveals that the low coordination at the zigzag boundaries induces a severe spatial symmetry breaking and localized charge accumulation, primarily accommodated by the in-plane $5d$ orbitals. As the ribbon width scales up, this edge-induced influence is progressively screened, and the spatial delocalization of states within the expanding central domain allows the system to recover the pristine metallic continuum of the infinite 2D sheet. Finally, the high electronic activity observed at the boundary sites suggests that these nanoribbons can serve as more reactive platforms than their central regions. Taken together, these findings provide a comprehensive framework for engineering the electronic structures of goldene derivatives, underscoring their potential for next-generation nanoelectronic devices, spintronics, and efficient catalysis.

\section{Acknowledgements}
L.A.R.J. acknowledges the financial support from FAP-DF grants 00193.00001808/2022-71 and $00193-00001857/2023-95$, FAPDF-PRONEM grant 00193.00001247/2021-20, PDPG-FAPDF-CAPES Centro-Oeste 00193-00000867/2024-94, and CNPq grants $350176/2022-1$ and $167745/2023-9$. A.C.D. acknowledges financial support from FAP-DF grants 00193-00001817/2023-43 and 00193-00002073/2023-84, CNPq grants 408144/2022-0, 305174/2023-1, 444069/2024-0, and 444431/2024-1. K.A.L.L. acknowledges the Center for Computational Engineering \& Sciences (CCES) at Unicamp for financial support through the FAPESP/CEPID Grant 2013/08293-7. E.C.G acknowledges the support of the Brazilian agency CNPq (Process No. 309832/2023-3). E.C.G acknowledges the support from Funda\c c\~ao de Amparo \`a Pesquisa do Estado do Piauí (FAPEPI) and CNPq through the PRONEM program. The authors thank the Laborat\'orio de Simula\c c\~ao Computacional Caju\'ina (LSCC) at Universidade Federal do Piau\'i for computational support. The authors also acknowledge computational support from Centro Nacional de Processamento de Alto Desempenho at Ceará (CENAPAD-UFC) and São Paulo (CENAPAD-SP). D. S. G. acknowlegdes support from INEO/CNPq and FAPESP
(grant 2025/27044-5).

\bibliographystyle{edu_style2}
\bibliography{Goldene_nanoribbons}

\clearpage 

\onecolumngrid 

\setcounter{page}{1} 
\setcounter{equation}{0} 
\setcounter{figure}{0} 
\setcounter{table}{0} 

\renewcommand{\theequation}{S\arabic{equation}}
\renewcommand{\thefigure}{S\arabic{figure}}
\renewcommand{\thetable}{S\arabic{table}}
\renewcommand{\thepage}{S\arabic{page}} 

\begin{center}
    \textbf{\large Supplementary Information for: \\ [0.5em] Electronic and structural properties of goldene nanoribbons}
    
    \vskip 0.5cm
    Fabricio et al. 
\end{center}
\vskip 1cm

\section{AGOLDNR bands}

\begin{figure}[h!]
\begin{center}
  \includegraphics[width=0.9\textwidth]{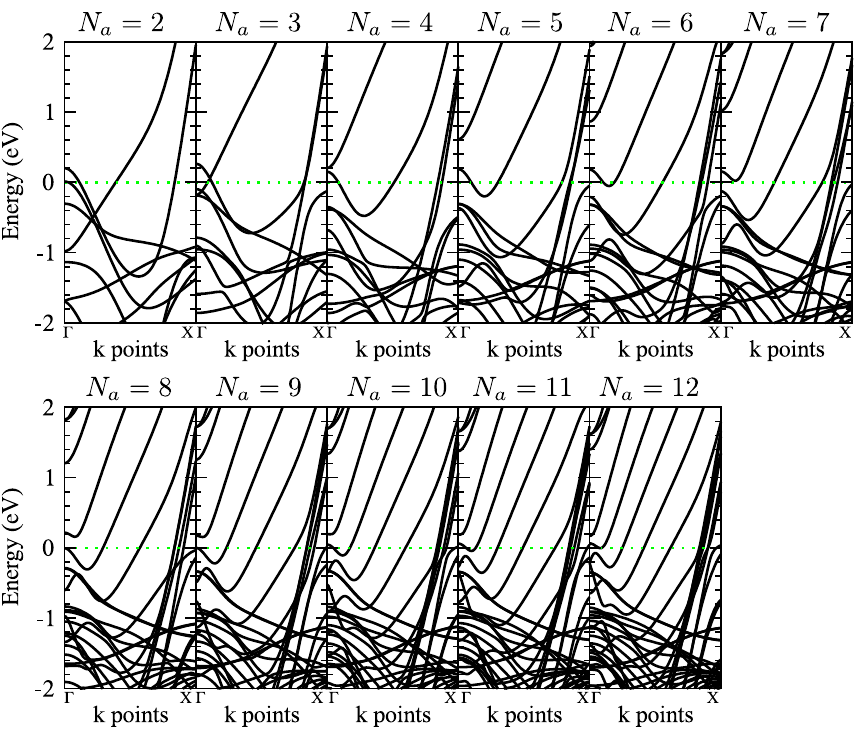}
  \caption{\footnotesize Electronic band structures of armchair goldene nanoribbons (AGOLDNRs) as a function of the ribbon width, ranging from $N_{a} = 2$ to $N_{a} = 12$. The Fermi level ($E_F$) is shifted to 0 eV and indicated by the green dotted line.}
    \label{fig:si_fig1}
  \end{center}
\end{figure}

\newpage

\section{ZGOLDNR bands}

\begin{figure}[h!]
\begin{center}
  \includegraphics[width=0.8\textwidth]{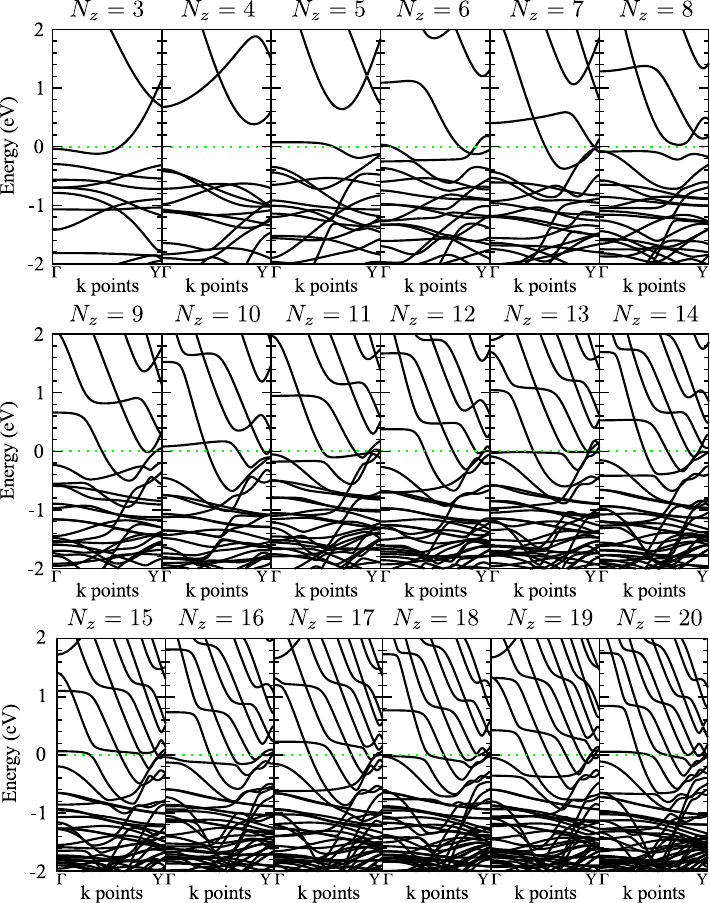}
  \caption{\footnotesize Electronic band structures of zigzag goldene nanoribbons (ZGOLDNRs) as a function of the ribbon width, ranging from $N_{z} = 3$ to $N_{z} = 20$. The Fermi level ($E_F$) is shifted to 0 eV and indicated by the green dotted line. Notably, a pronounced band-braiding effect near $E_F$ emerges and becomes clearly resolved in the wider ribbon regime ($N_{z} > 10$), showing that long-range orbital rehybridization and symmetry breaking driven by the low-coordination-number ($CN=3$) boundary sites manifest robustly as the system expands.}
    \label{fig:si_fig2}
  \end{center}
\end{figure}

\section{ZGOLDNR bands in FM configurations}

\begin{figure}[h!]
\begin{center}
  \includegraphics[width=0.85\textwidth]{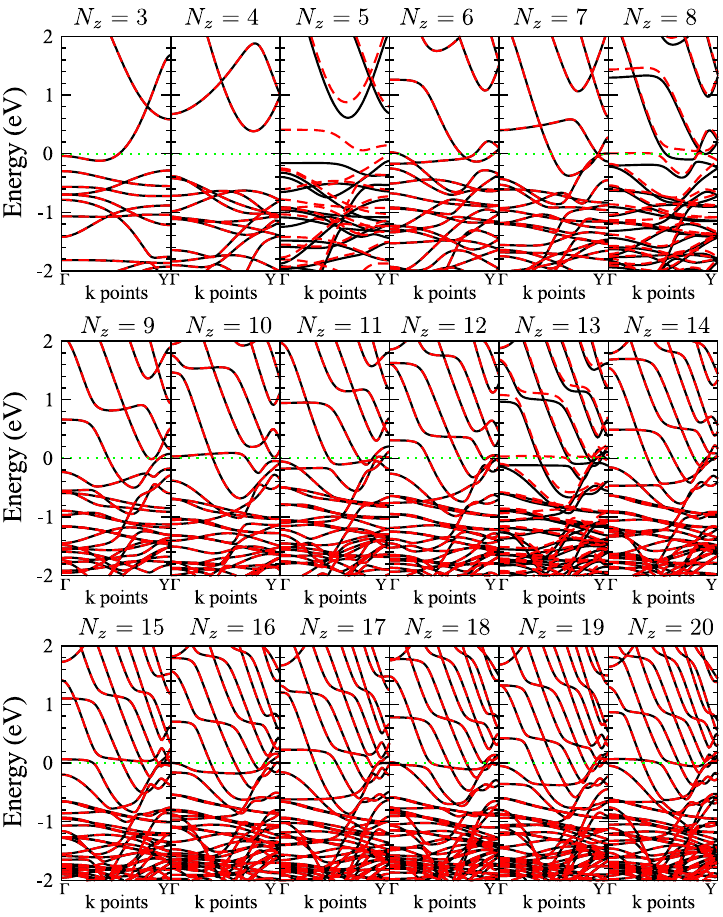}
  \caption{\footnotesize Electronic band structures of zigzag goldene nanoribbons (ZGOLDNRs) as a function of the ribbon width, ranging from $N_{z} = 3$ to $N_{z} = 20$ in the ferromagnetic (FM) configuration. The Fermi level ($E_F$) is shifted to 0 eV and indicated by the green dotted line.}
    \label{fig:si_fig3}
  \end{center}
\end{figure}

\end{document}